\documentclass[aps,prl,showpacs,preprint,amsmath,amssymb,superscriptaddress]{revtex4}

\usepackage{graphicx}
\begin{document}
\bibliographystyle{prsty}
\title{Indication of antiferromagnetic interaction between paramagnetic Co ions in the diluted magnetic semiconductor Zn$_{1-x}$Co$_{x}$O}

\author{M.~Kobayashi}
\altaffiliation{Present address: Department of Applied Chemistry, 
School of Engineering, University of Tokyo, 
7-3-1 Hongo, Bunkyo-ku, Tokyo 113-8656, Japan}
\affiliation{Department of Physics, University of Tokyo, 
7-3-1 Hongo, Bunkyo-ku, Tokyo 113-0033, Japan}
\author{Y.~Ishida}
\altaffiliation{Present address: RIKEN, SPring-8 Center, Sayo-cho, Hyogo 679-5148, Japan}
\affiliation{Department of Physics, University of Tokyo, 
7-3-1 Hongo, Bunkyo-ku, Tokyo 113-0033, Japan}
\author{J.~I.~Hwang}
\affiliation{Department of Physics, University of Tokyo, 
7-3-1 Hongo, Bunkyo-ku, Tokyo 113-0033, Japan}
\author{Y.~Osafune}
\affiliation{Department of Physics, University of Tokyo, 
7-3-1 Hongo, Bunkyo-ku, Tokyo 113-0033, Japan}
\author{A.~Fujimori}
\affiliation{Department of Physics, University of Tokyo, 
7-3-1 Hongo, Bunkyo-ku, Tokyo 113-0033, Japan}
\author{Y.~Takeda}
\affiliation{Synchrotron Radiation Research Unit, 
Japan Atomic Energy Agency, Sayo-gun, Hyogo 679-5148, Japan}
\author{T.~Okane}
\affiliation{Synchrotron Radiation Research Unit, 
Japan Atomic Energy Agency, Sayo-gun, Hyogo 679-5148, Japan}
\author{Y.~Saitoh}
\affiliation{Synchrotron Radiation Research Unit, 
Japan Atomic Energy Agency, Sayo-gun, Hyogo 679-5148, Japan}
\author{K.~Kobayashi}
\affiliation{National Institute for Materials Science (NIMS), SPring-8, 1-1-1 Kouto, Sayo-cho, Sayo-gun, Hyogo 679-5148, Japan}
\author{H.~Saeki}
\affiliation{Institute of Scientific and Industrial Research, Osaka University, 
Ibaraki, Osaka 567-0047, Japan}
\author{T.~Kawai}
\affiliation{Institute of Scientific and Industrial Research, Osaka University, 
Ibaraki, Osaka 567-0047, Japan}
\author{H.~Tabata}
\affiliation{Department of Electronic Engineering, School of Engineering, University of Tokyo, 
7-3-1 Hongo, Bunkyo-ku, Tokyo 113-8656, Japan} 
\affiliation{Department of Bioengineering, School of Engineering, University of Tokyo, 7-3-1 Hongo, Bunkyo-ku, Tokyo 113-8656, Japan}
\date{\today}

\begin{abstract}
The magnetic properties of Zn$_{1-x}$Co$_x$O ($x=0.07$ and 0.10) thin films, which were homo-epitaxially grown on a ZnO(0001) substrates with varying relatively high oxygen pressure, have been investigated using x-ray magnetic circular dichroism (XMCD) at Co $2p$ core-level absorption edge. 
The line shapes of the absorption spectra are the same in all the films and indicate that the Co$^{2+}$ ions substitute for the Zn sites. 
The magnetic-field and temperature dependences of the XMCD intensity are consistent with the magnetization measurements, indicating that except for Co there are no additional sources for the magnetic moment, and demonstrate the coexistence of paramagnetic and ferromagnetic components in the homo-epitaxial Zn$_{1-x}$Co$_{x}$O thin films, in contrast to the ferromagnetism in the hetero-epitaxial Zn$_{1-x}$Co$_{x}$O films studied previously. 
The analysis of the XMCD intensities using the Curie-Weiss law reveals the presence of antiferromagnetic interaction between the paramagnetic Co ions. 
Missing XMCD intensities and magnetization signals indicate that most of Co ions are non-magnetic probably because they are strongly coupled antiferromagnetically with each other. 
Annealing in a high vacuum reduces both the paramagnetic and ferromagnetic signals. We attribute the reductions to thermal diffusion and aggregation of Co ions with antiferromagnetic nanoclusters in Zn$_{1-x}$Co$_{x}$O. 
\end{abstract}


\maketitle
\section{Introduction}
Ferromagnetic diluted magnetic semiconductors (DMS's) are key materials for future ``spin electronics'' or ``spintronics'', in which one manipulates both the spin and charge degrees of freedom of electrons, because interaction between the spins of host $sp$-band electrons and the magnetic moments of doped magnetic ions enables us to control the spin degrees of freedom of electrons \cite{Science_98_Ohno, JSAP_02_Ohno}. 
In fact, the ferromagnetism of the III-V DMS's Ga$_{1-x}$Mn$_{x}$As and In$_{1-x}$Mn$_{x}$As arises from carrier-mediated interaction between Mn spins \cite{RMP_06_Jungwirth}, and new functional devices utilizing both magnetism and electrical conductivity have been fabricated based on these materials, e.g., spin-dependent circular light-emitting diodes \cite{Nature_99_Ohno} and field-effect transistors controlling ferromagnetism \cite{Nature_00_Ohno}. 
New control methods of magnetization have also been demonstrated, such as electrical manipulation of magnetization reversal \cite{Science_03_Chiba} and current-induced domain-wall switching \cite{Nature_04_Yamanouchi}. 
For practical applications of such devices, ferromagnetic DMS's having Curie temperatures ($T_\mathrm{C}$'s) above room temperature are necessary. 
Recently, oxide-based DMS's have attracted much attention first motivated by the theoretical predictions of high $T_\mathrm{C}$'s in wide-gap semiconductors \cite{Science_00_Dietl, SST_02_Sato} followed by various reports of ferromagnetism at room temperature \cite{SST_04_Pearton}.

Ever since the discovery of ferromagnetism in Zn$_{1-x}$Co$_{x}$O thin films \cite{APL_01_Ueda}, ZnO-based DMS's have been investigated intensively because the host material ZnO has potential for optical applications due to the wide band gap of $3.4$ eV and the large exciton energy of $60$ meV. In particular, Zn$_{1-x}$Co$_{x}$O is one of the most investigated DMS's, and there have been many reports about the magnetic properties of Zn$_{1-x}$Co$_{x}$O. 
Schwartz and Gamelin \cite{AdvMater_04_Schwartz} have reported reversible switching of room-temperature ferromagnetism in Zn$_{1-x}$Co$_{x}$O by introducing and removing interstitial Zn defects. Reproducibly, exposure of Zn$_{1-x}$Co$_{x}$O thin film to Zn vapor creates interstitial Zn defects and induces room-temperature ferromagnetism, and annealing the thin film in air quenches the ferromagnetism due to elimination of the interstitial Zn by oxidation. 
Neal {\it et al}. \cite{PRL_06_Neal} have observed magnetic circular dichroism (MCD) signals with open hysteresis loop at the ZnO band edge of Co-doped ZnO thin films at room temperature, implying a splitting of the conduction band of ZnO through interaction with the magnetic moments of the Co ions. These observations rule out Co-cluster segregations or secondary phase formation as the origin of the ferromagnetism. 
In addition, there are some reports about the observation of anomalous Hall effects in Zn$_{1-x}$Co$_{x}$O \cite{APL_08_Hsu, JAP_08_Yang}, providing experimental evidence for carrier-mediated ferromagnetism in Zn$_{1-x}$Co$_{x}$O. 
Because the preparation of Zn$_{1-x}$Co$_{x}$O thin films has not been highly reproducible yet and secondary phases, defects, and carriers likely affect the magnetic behavior of Zn$_{1-x}$Co$_{x}$O, 
systematic measurements on well-defined Zn$_{1-x}$Co$_{x}$O thin films are necessary to understand the origin of the ferromagnetism in Zn$_{1-x}$Co$_{x}$O.

Recently, Matsui and Tabata \cite{PRB_07_Matsui} have succeeded in homo-epitaxial growth of Co-doped ZnO on a ZnO substrate. The homo-epitaxial growth has enabled us to obtain more reproducible high-quality Zn$_{1-x}$Co$_{x}$O thin films, providing correlation between the growth parameters and the physical properties. The magnetization measurements of the homo-epitaxial Zn$_{1-x}$Co$_{x}$O thin films indicated that the spontaneous magnetization ($M_s$) depends on electron-carrier concentration ($n$) controlled by oxygen pressure during deposition. 
In the Zn$_{1-x}$Co$_x$O thin films, $M_s$ increases with increasing $n$, consistent with the result of anomalous Hall effect measurements \cite{APL_08_Hsu, JAP_08_Yang}.

Since the ferromagnetism in Zn$_{1-x}$Co$_{x}$O has been debated between researchers using different techniques, element-specific magnetization measurements provide useful information in order to address the magnetic properties of Zn$_{1-x}$Co$_{x}$O. 
X-ray magnetic circular dichroism (XMCD) is one of the most powerful tools to investigate the magnetic properties of materials. XMCD is defined as the difference between absorption spectra for different circular polarized x rays. 
In the previous XMCD measurements on a hetero-epitaxial Zn$_{1-x}$Co$_{x}$O thin film grown on Al$_2$O$_3$(0001) substrates \cite{PRB_05_Kobayashi}, we have concluded that the ferromagnetism comes from Co ions substituting the Zn site in ZnO, and proposed that non-ferromagnetic Co ions are strongly coupled antiferromagnetically with each other from the linear magnetic-field ($H$) dependence and temperature ($T$) independence of XMCD. Sati {\it et al}. \cite{PRL_07_Sati} have reported antiferromagnetic interaction in single crystalline Zn$_{1-x}$Co$_{x}$O thin films with low Co concentration ($x = 0.003 - 0.005$) from magnetic and EPR measurements. 
On the other hand, there are some reports that the Co $3d$ electrons are not the origins of the ferromagnetism \cite{PRB_07_Barla} or that metallic Co atoms contribute to the ferromagnetism \cite{APL_08_Rode}. 
Since the magnetic behavior of Zn$_{1-x}$Co$_{x}$O may be affected by extrinsic factors such as secondary phases and defects, systematic XMCD measurements on well-defined Zn$_{1-x}$Co$_{x}$O thin films will provide us with useful insight into the magnetism.

Recently, annealing effects on the ferromagnetism in Zn$_{1-x}$Co$_x$O have been discussed in relation to oxygen vacancies and/or Zn interstitials \cite{PRL_06_Kittilstved, AdvMater_07_MacManus-Driscoll}. 
Annealing of Zn$_{1-x}$Co$_{x}$O in reducing atmosphere such as in high vacuum, which may produce oxygen vacancies, have been reported to enhance the ferromagnetism, while annealing in oxidizing atmosphere such as air decreases the magnetic moments of Zn$_{1-x}$Co$_{x}$O \cite{APL_06_Hsu, Nanotech_06_Wu, JPCM_07_Sudakar, APL_07_Zhao}. In addition, we note that ferromagnetism has been reported in some oxide thin films without magnetic ions \cite{PRB_06_Sundaresan, MMM_07_Hong, JPCM_08_Ghoshal}, probably due to defects such as oxygen vacancies. 
It is therefore important to investigate annealing effects and/or the role of oxygen vacancies for the understanding of the ferromagnetism in Zn$_{1-x}$Co$_{x}$O. 

In the present work, we report on the results of $H$- and $T$-dependent XMCD measurements on homo-epitaxial Zn$_{1-x}$Co$_{x}$O thin films which show coexisting paramagnetic and ferromagnetic properties. 
Analysis using the Curie-Weiss law suggests that antiferromagnetic correlation between the Co ions suppress the paramagnetic behavior. We have also found that the XMCD intensity decreases by high-vacuum annealing. Based on the experimental findings, a possible origin of the ferromagnetism in the Zn$_{1-x}$Co$_{x}$O thin films shall be discussed.

\section{Experimental}
Zn$_{1-x}$Co$_x$O thin films were homo-epitaxially grown on Zn-terminated ZnO(0001) substrates by the pulsed laser deposition technique with varying oxygen pressure ($P_{\mathrm{O}2}$) from $1 \times 10^{-4}$ to 10$^{-2}$ Pa. Four thin films were prepared: a $x$=0.07 film deposited at $P_{\mathrm{O}2}$=10$^{-4}$ Pa, and $x$=0.10 films deposited at $P_{\mathrm{O}2}$=10$^{-2}$, 10$^{-3}$, and 10$^{-4}$ Pa. 
During the deposition, the substrate temperature was kept at $400$ $^{\circ}$C. The thickness of each film was $50 - 150$ nm. X-ray diffraction confirmed that the thin films had the wurtzite structure and no secondary phase was observed. 
The magnetization of those thin films was measured using a superconducting quantum interference device (SQUID) magnetometer (Quantum Design, Co. Ltd). 
The present homo-epitaxial films are expected to have higher crystal quality \cite{PRB_07_Matsui} than the hetero-epitaxial one studied in the previous XMCD measurements \cite{PRB_05_Kobayashi}. However, the homo-epitaxial thin films have lower $M_s$ than the hetero-epitaxial ones because of the high $P_\mathrm{O2}$ values \cite{PRB_07_Matsui}.

X-ray absorption spectroscopy (XAS) and XMCD measurements were performed at the helical undulator beam line BL23SU of SPring-8 \cite{JSR_98_Yokoya, AIP_04_Okamoto}. The monochromator resolution was $E/{\Delta}E \textgreater 10,000$. Absorption spectra of right-handed (${\mu}^+$) and left-handed (${\mu}^-$) circularly polarized x rays were obtained by reversing photon helicity at each photon energy in the total-electron-yield mode. The $\mu^{+}$ and $\mu^{-}$ spectra were averaged between spectra for the positive and negative applied magnetic fields so as to eliminate spurious signals. The degree of circular polarization of x-rays was over 90 \%. 
External magnetic fields were applied perpendicular to the sample surface. Circularly-polarized x-ray absorption spectra under each experimental conditions have been normalized to the maximum height of the Co $2p$ XAS [$(\mu^+ + \mu^-)/2$] spectrum. Backgrounds of the XAS spectra at the Co $2p$ edge were assumed to be hyperbolic tangent functions. 
Annealing of the $P_\mathrm{O2}$=10$^{-2}$ Pa sample was performed at 700 $^{\circ}$C and $\sim$10$^{-7}$ Pa for 5 min. For surface cleaning, Ar$^+$ ion sputtering and subsequent 250 $^{\circ}$C annealing (well below the deposition temperatures) were performed in a high vacuum of 10$^{-7}$ Pa.

\section{Results and discussion}
\begin{figure}[p!]
\begin{center}
\includegraphics[width=14cm]{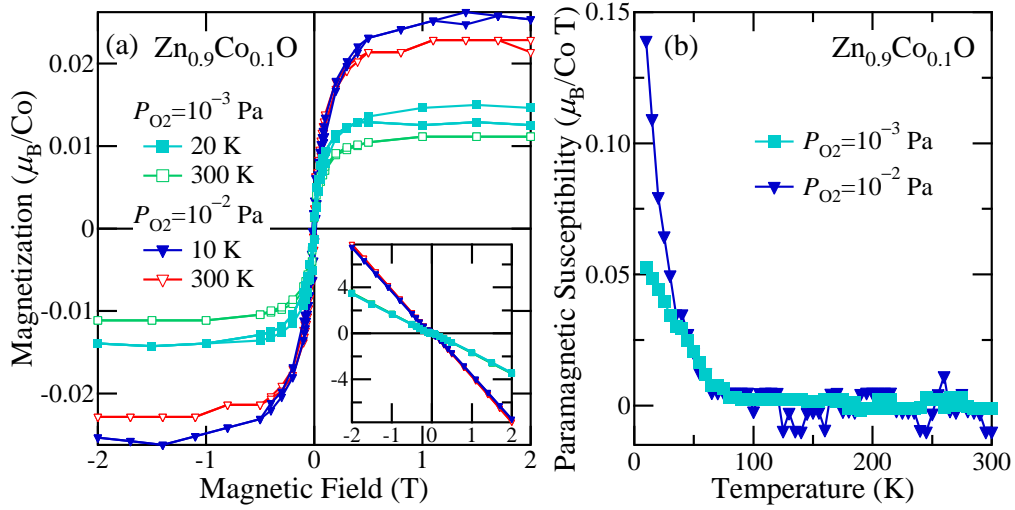}
\caption{(Color online) Magnetic properties of Zn$_{0.9}$Co$_{0.1}$O thin films measured by a SQUID magnetometer. (a) Magnetic-field ($H$) dependence of the magnetization. The $H$-linear components deduced from high $H$ data have been subtracted. The inset shows raw data containing the diamagnetic signals from the substrates. 
(b) Temperature ($T$) dependence of the paramagnetic component of the magnetization represented in the form of magnetic susceptibility, where the $T$-independent component deduced from high-$T$ data have been subtracted. 
}
\label{SQUID}
\end{center}
\end{figure}

Figure~\ref{SQUID} shows the magnetic-field ($H$) and temperature ($T$) dependences of the magnetization ($M$) measured using a SQUID magnetometer. 
Although the raw magnetization data decreases with $H$ due to the diamagnetic contribution from the substrate as shown in the inset of Fig.~\ref{SQUID}(a), there are small but finite ferromagnetic signals. 
By subtracting the linear component at high magnetic fields from the raw data, the ferromagnetic hysteresis loops have been extracted at 10 and 300 K, as shown in the main panel of Fig.~\ref{SQUID}(a). The hysteresis loop at 300 K implies that $T_\mathrm{C}$ of the ferromagneitc component is above room temperature. 
Here, the coercive fields are found to be 100-200 Oe. The $M_s$ of the films was of the order of $10^{-2}$ $\mu_\mathrm{B}$/Co, indicating that less than 1 \% of the Co ions participate in the ferromagnetism if the Co ions is in the high spin Co$^{2+}$ state. The $M_s$ slightly decreases with increasing temperature. 
The $M_s$'s of the present homo-epitaxial films are as small as $5-10$ \% of that of the previous hetero-epitaxial one because of the high $P_\mathrm{O2}$, although the present Zn$_{1-x}$Co$_x$O thin film has higher crystal quality than the previous one. 
Assuming that the diamagnetic and ferromagnetic responses hardly change with temperature, the paramagnetic susceptibilities have been obtained by subtracting the slopes of the $M$-$H$ curves taken at different $T$'s, as shown in Fig.~\ref{SQUID}(b). One can see that the paramagnetic signals increase with decreasing $T$ in the low $T$ region. 
The observations thus suggest the presence of both the paramagnetic and ferromagnetic components in the present Zn$_{1-x}$Co$_{x}$O thin films.

\begin{figure}[p!]
\begin{center}
\includegraphics[width=14cm]{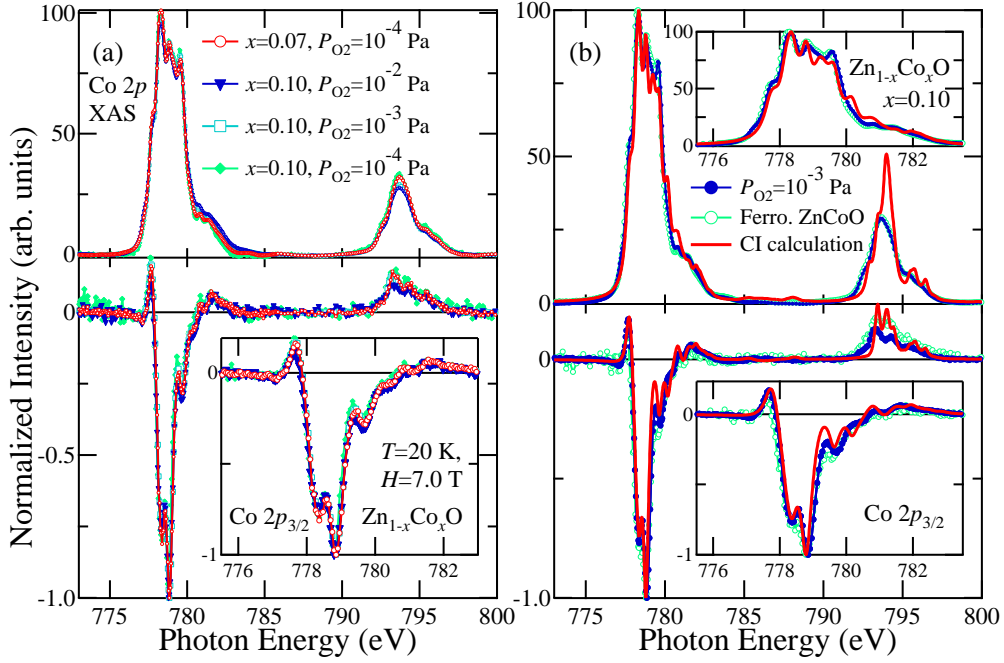}
\caption{(Color online) Co $2p$ core-level absorption spectra of Zn$_{1-x}$Co$_{x}$O thin films measured at $H=7.0$ T and $T=20$ K. 
(a) Comparison of the XAS (upper) and XMCD (lower) spectra between the samples measured at $H=7.0$ T and $T=20$ K. 
(b) Comparison of the line shapes of the present spectra with that of the ferromagnetic sample and the calculated spectrum reported in the previous study \cite{PRB_05_Kobayashi}. All the XAS and XMCD spectra have been normalized to the positive and negative peak heights, respectively. 
The insets show expanded plots at the Co $2p_{3/2}$ edge. 
}
\label{Co3dStates}
\end{center}
\end{figure}

Figure~\ref{Co3dStates} shows the Co $2p$ XAS and XMCD spectra of the samples measured at $H=7.0$ T and $T=20$ K. The XAS and XMCD spectra have been normalized to the positive and negative peak heights, respectively. We have observed clear XMCD signals for all the thin films. As shown in Fig.~\ref{Co3dStates}(a), the line shapes of the XAS and XMCD spectra are identical between these films, suggesting that the doped Co ions have almost the same electronic structure. 
The XAS and XMCD spectra are compared with these of the previous report on the hetero-epitaxial Zn$_{1-x}$Co$_{x}$O thin film and the cluster-model calculation for the Co$^{2+}$ ion under tetrahedral crystal field \cite{PRB_05_Kobayashi}, as shown in Fig.~\ref{Co3dStates}(b). The line shapes of the XAS and XMCD spectra well agree with these of the previous ones and are similar to those of the calculation for the Co$^{2+}$ ion tetrahedrally coordinated by oxygen atoms, indicating that the Co ions substitute for the Zn sites.

\begin{figure}[p!]
\begin{center}
\includegraphics[width=14cm]{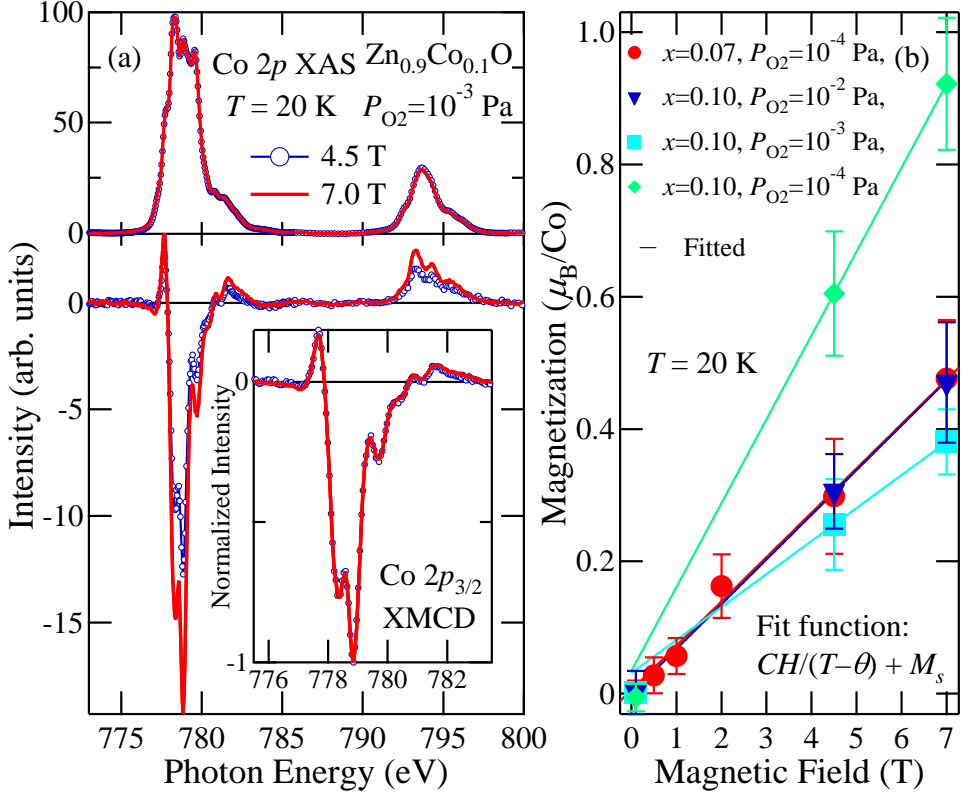}
\caption{(Color online) Magnetic-field dependence of the XMCD spectra of Zn$_{1-x}$Co$_{x}$O thin films. 
(a) XMCD spectra of the $x=0.10$, $P_\mathrm{O2} = 10^{-3}$ sample at $T=20$ K and at various magnetic fields. The inset shows the XMCD spectra normalized to the negative peak height at $h\nu=778.5$ eV. 
(b) $H$ dependence of the total magnetization ($M_\mathrm{tot}$) estimated using the XMCD sum rules at $T=20$ K. Solid curves are the results of fitting to the Curie-Weiss law, $CH/(T-\theta) + M_s$. 
}
\label{H-depXMCD}
\end{center}
\end{figure}

\begin{figure}[p!]
\begin{center}
\includegraphics[width=14cm]{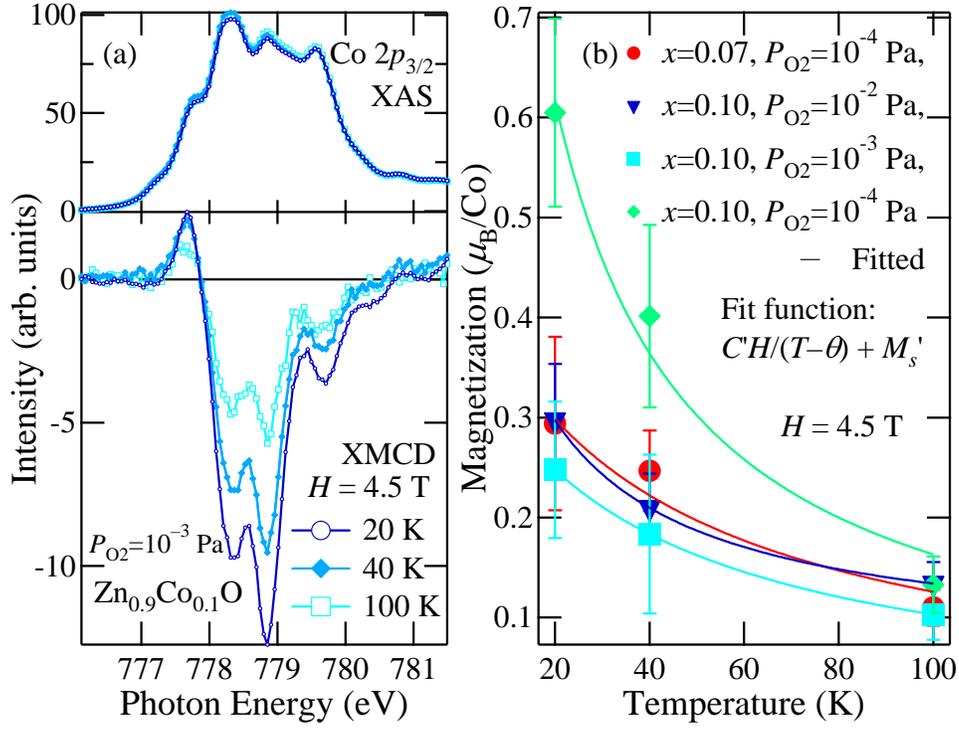}
\caption{(Color online) Temperature dependence of the XMCD spectra of Zn$_{1-x}$Co$_{x}$O thin films. 
(a) XMCD spectra of the $x=0.10$, $P_\mathrm{O2}$=10$^{-3}$ Pa sample at $H=4.5$ T and at various temperatures. 
(b) $T$ dependence of the total magnetic moment $M_\mathrm{tot}$ estimated using the XMCD sum rules at $H=4.5$ T. Solid curves are the results of fitting to the Curie-Weiss law, $C'H/(T-\theta) + M_s'$. 
}
\label{T-depXMCD}
\end{center}
\end{figure}

In order to understand the magnetic properties of the present Zn$_{1-x}$Co$_{x}$O thin films, we have studied the $H$ and $T$ dependences of the XMCD spectra. Figure~\ref{H-depXMCD} shows the $H$ dependence of the XMCD spectra of the Zn$_{0.9}$Co$_{0.1}$O thin film deposited at $P_{\mathrm{O}2}$=10$^{-3}$ Pa. The line shape of the XMCD spectrum is unchanged with the strength of applied magnetic field, as shown in the inset of Fig.~\ref{H-depXMCD}(a). 
The total magnetization $M_\mathrm{tot} = M_\mathrm{spin} + M_\mathrm{orb}$ estimated from the XMCD sum rules of these thin films are plotted as a function of $H$ in Fig.~\ref{H-depXMCD}(b). Here, $M_\mathrm{spin}$ and $M_\mathrm{orb}$ are spin and orbital magnetic moments, respectively. In all the films, $M_\mathrm{tot}$ linearly increases with $H$, and $M_s$ obtained by extrapolating $M_\mathrm{tot}$ to $H = 0$ ($M_s^{\mathrm{XMCD}}$) is very small, quaritatively consistent with the magnetization measurements. 
The observations suggest that in the present Zn$_{1-x}$Co$_{x}$O films the paramagnetic signals are predominant compared with the ferromagnetic ones unlike the previous sample \cite{PRB_05_Kobayashi}. 
Figure~\ref{T-depXMCD} shows the $T$ dependence of the XMCD spectra of the Zn$_{0.9}$Co$_{0.1}$O thin film deposited at $P_{\mathrm{O}2}$=10$^{-3}$ Pa. The line shapes of the XMCD spectra were independent of temperature although the XMCD intensity decreased with increasing $T$. The $M_s^{\mathrm{XMCD}}$'s of all the samples as a function of $T$ are summarized in Fig.~\ref{T-depXMCD}(b).

Although the XMCD intensities were different between these films, all the XMCD spectra of the thin films showed similar $H$ and $T$ dependences. 
According to the relationship between the $M_s$ and $n$ of the homo-epitaxial Zn$_{1-x}$Co$_x$O thin films \cite{PRB_07_Matsui}, the effective $n$'s of all the samples studied here are estimated to be less than 10$^{18}$ cm$^{-3}$. 
The observations that the XMCD intensity increases and decreases with increasing $H$ and $T$, respectively, indicate a Curie-Weiss paramagnetic behavior, in spite of almost the same electronic structure of the Co ions as that of the ferromagnetic hetero-epitaxial Zn$_{1-x}$Co$_{x}$O thin film [see Fig.~\ref{Co3dStates}(b)], which did not show a Curie-Weiss paramagnetic component. 
The results imply that although the Co ions substitute for the Zn sites, the Co ions become paramagnetic or ferromagnetic in Zn$_{1-x}$Co$_{x}$O depending on $n$ and the crystallinity. 
Considering that Co $2p$ XMCD reflects only the local electronic states of the Co ion, it is likely that the different magnetic responses of the Zn$_{1-x}$Co$_{x}$O films are originated from different nearest and next-nearest neighbor interaction with neighboring cations, defects, and/or carrier concentration. 
We shall discuss about the differences in the magnetic properties below.

Let us quantitatively discuss the paramagnetic behavior of the Co ions by applying XMCD sum rules to the spectra \cite{PRL_92_Thole, PRL_93_Carra}. 
The $H$ and $T$ dependences of $M_\mathrm{tot}$ have been fitted to the formula
\begin{equation}
\langle M_\mathrm{tot} \rangle = \frac{CH}{T - \theta} + M_s, 
\label{CW-law}
\end{equation}
where $C$ is the Curie constant and $\theta$ is the Weiss temperature. The second term $M_s$ in Eq.~(\ref{CW-law}) is contributions from the ferromagnetic component. 
Results of the fit are shown in Figs.~\ref{H-depXMCD}(b) and \ref{T-depXMCD}(b), indicating that Eq.~(\ref{CW-law}) well reproduces both the $H$ and $T$ dependences of $M_\mathrm{tot}$. The fitted parameters are listed in Table~\ref{CW_analysis}. 
It should be noted that the obtained $M_s$'s well agree with those deduced from the magnetization measurements as shown in Table~\ref{CW_analysis}, providing experimental evidence that we have probed the intrinsic bulk magnetic properties of the Zn$_{1-x}$Co$_x$O thin films. In addition, the agreement suggests that magnetic moment of the substituted Co ions is predominant source of the ferromagnetic responses of the samples expected to be in the low $n$ region. 
The $C$ values estimated from the Curie-Weiss fit are close to that of the free Co$^{2+}$ ion with quenched orbital moment of $\sim$3.36 $\mu_\mathrm{B}$K/T, except for the $P_{\mathrm{O}2}$=10$^{-2}$ Pa thin film. 
The Weiss temperature $\theta$ of all the samples are negative, consistent with the report of antiferromagnetic interaction in paramagnetic Zn$_{1-x}$Co$_{x}$O thin films \cite{PRL_07_Sati}.

\begin{figure}[p!]
\begin{center}
\includegraphics[width=14cm]{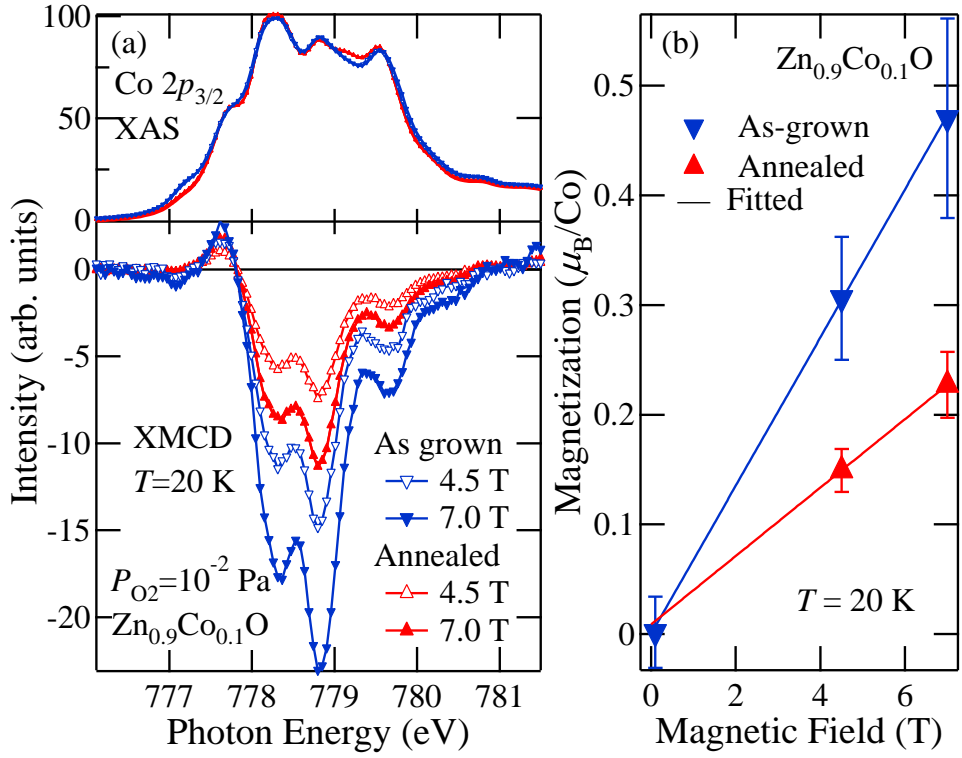}
\caption{(Color online) Annealing effects on the XMCD spectra of the Zn$_{0.9}$Co$_{0.1}$O thin film deposited at $P_\mathrm{O2} = 10^{-2}$ Pa. 
(a) Comparison of the $H$ dependent XMCD spectra between the as-grown and annealed samples. 
(b) $H$ dependence of the magnetization of the same samples deduced from the XMCD spectra. The $H$ dependences have been fitted to linear functions $\chi H + M_s$. 
}
\label{Annealing}
\end{center}
\end{figure}

\begin{figure}[p!]
\begin{center}
\includegraphics[width=14cm]{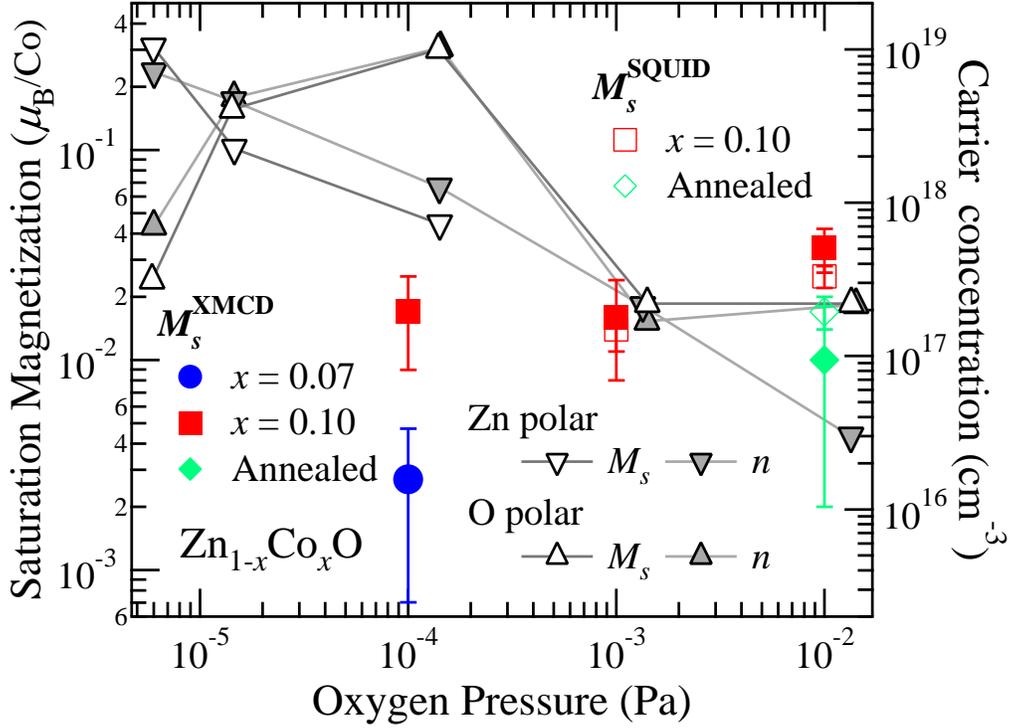}
\caption{(Color online) $P_\mathrm{O2}$ dependences of saturation magnetization $M_s$ and carrier concentration $n$ in homo-epitaxial Zn$_{1-x}$Co$_x$O thin films. 
The values of O- and Zn-polar Zn$_{1-x}$Co$_x$O are taken from Ref.~\cite{PRB_07_Matsui}. Here, all the experimental values of the present study are plotted based on $M_s$ and $P_{\mathrm{O2}}$. 
}
\label{MNvsP}
\end{center}
\end{figure}

In order to see the effects of vacuum annealing on the electronic and magnetic properties of the Co ion, we have measured the Co $2p$ XMCD spectra of the $P_{\mathrm{O}2}$=10$^{-2}$ Pa film after vacuum annealing, too. 
Figure~\ref{Annealing}(a) shows the Co $2p_{3/2}$ XAS and XMCD spectra of the $P_{\mathrm{O}2}$=10$^{-2}$ Pa film before and after the annealing. The line shapes of the XAS and XMCD spectra are almost identical before and after the annealing, indicating that the local electronic structure of the Co ion remained unchanged by the annealing. However, the XMCD intensity of the annealed sample becomes about half of the value of the as-grown one. The $M_s$ of the annealed film has been estimated by fitting of the $H$ dependence of the $M_\mathrm{tot}$ to a linear function
\begin{equation}
\langle M_\mathrm{tot} \rangle = \chi H + M_s, 
\label{LinearFunction}
\end{equation}
as shown in Fig.~\ref{Annealing}(b), and the results of the fitting are listed in Table~\ref{EffectsOfAnnealing}. The slopes of the $M_\mathrm{tot}$ versus $H$ curve, namely, the susceptibility $\chi$ became half of the as-grown sample by the annealing. 
The value of $M_s$ of $1.0 \pm 0.8 \times 10^{-2}$ $\mu_\mathrm{B}$/Co became also somewhat smaller than that of the as-grown one. 
The observations indicate that both the paramagnetic and ferromagnetic components were suppressed by the annealing, in contrast to the reports on high-vacuum annealing on Zn$_{1-x}$Co$_{x}$O thin films that the ferromagnetism of Zn$_{1-x}$Co$_{x}$O thin films increased after annealing in a high vacuum \cite{Nanotech_06_Wu, JPCM_07_Sudakar, APL_07_Zhao}. 
The agreement of $M_s$ between the XMCD and SQUID measurements implies that the substituted Co ions are also predominantly responsible for the ferromagnetism even after the annealing process. 
Figure~\ref{MNvsP} shows the $P_\mathrm{O2}$ dependences of $M_s$ and $n$ in the homo-epitaxial Zn$_{1-x}$Co$_x$O thin films \cite{PRB_07_Matsui}. The $M_s$ increases with increasing $n$ in both the O- and Zn-polar Zn$_{1-x}$Co$_x$O. 
Taking into account the correlation between $M_s$ and $n$ that the increase in $n$ increases $M_s$ as show in Fig.~\ref{MNvsP}, we conclude that the annealing hardly created enough oxygen vacancies in the thin film and $n$ of the film remained less than 10$^{-19}$ cm$^{-3}$ \cite{PRB_07_Matsui, APL_08_Hsu, JAP_08_Yang} and that the suppressions of the paramagnetic and ferromagnetic magnetizations by the annealing are not originated from defect formation nor carrier doping.

\begin{figure}[p!]
\begin{center}
\includegraphics[width=14cm]{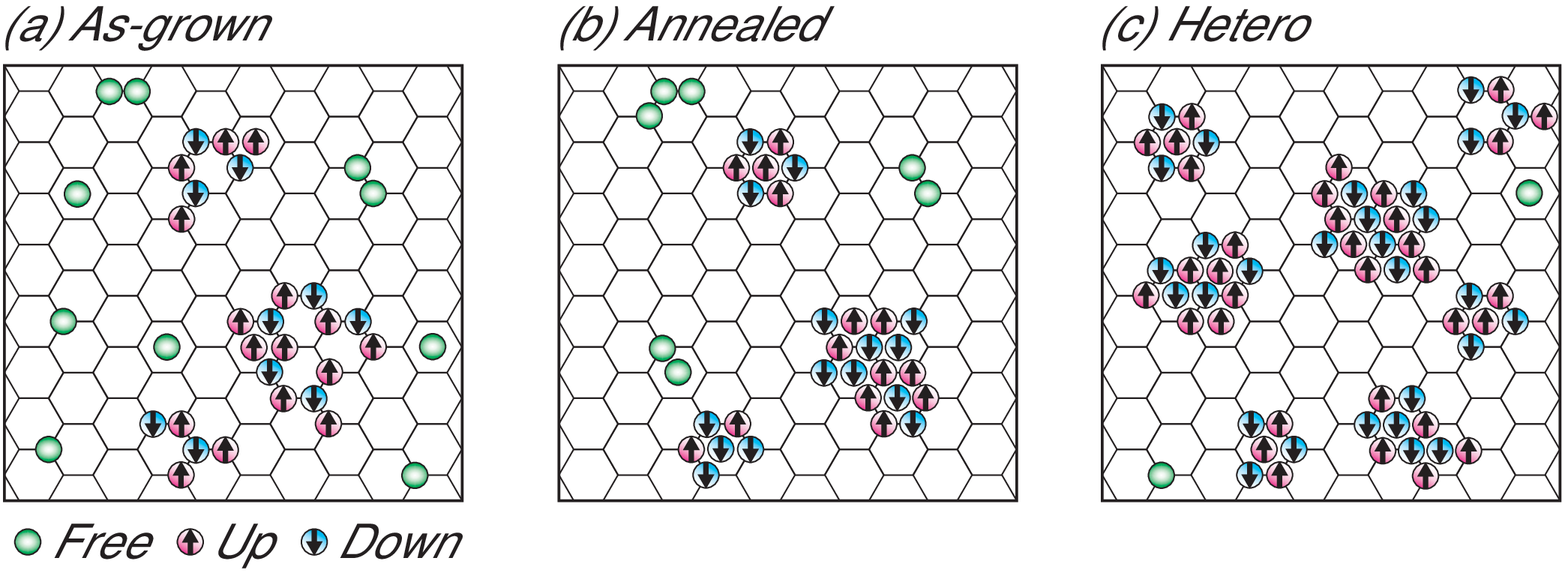}
\caption{(Color online) Schematic drawings of Co-ion distribution in Zn$_{1-x}$Co$_x$O. 
(a), (b) Co distribution of the as-grown and annealed homo-epitaxial Zn$_{1-x}$Co$_x$O thin films, respectively. 
(c) Hetero-epitaxial Zn$_{1-x}$Co$_x$O thin film. 
}
\label{CoAgg}
\end{center}
\end{figure}

Let us then discuss the effects of annealing on the magnetic properties of Zn$_{1-x}$Co$_{x}$O by considering the antiferromagnetic interaction between the Co ions. 
Recently, Dietl {\it et al.} \cite{PRB_07_Dietl} have observed that the blocking temperature of Zn$_{1-x}$Co$_{x}$O thin films, which show both ferromagnetism and paramagnetism, reaches room temperature, and have proposed that the ferromagnetism of Zn$_{1-x}$Co$_{x}$O results from antiferromagnetic nanoclusters of Co-rich wurtzite (Zn,Co)O created by spinodal decomposition, similar to the ferromagnetism in NiO nanoparticles \cite{PRL_97_Kodama, PRB_05_Winkler}. 
In this model, the finite moments are originated from uncompensated spins on the surfaces of the nanoclusters and the average moments shall decrease with increasing nanocrystal size. 
Similarly, for bulk Zn$_{1-x}$Co$_x$O, it has been reported that the dominant interaction between the Co ions is antiferromagnetic and the magnetic propertiy can be explained by asuuming the existence of antiferromagnetic Co clusters \cite{PRB_05_Lawes}. 
Based on the above picture, the annealing effects may be understood as thermal diffusion and aggregation of isolated paramagnetic Co ions into antiferromagnetic nanoclusters, as follows. 
In the as-grown sample, there exist isolated Co ions and a small number of nanoclusters, which contribute to the Curie-Weiss paramagnetism and the weakly ferromagnetic antiferromagnetism, respectively. If the Co ions tend to aggregate in ZnO, thermal diffusion of Co ions by the annealing may lead the decrease of the number of isolated Co ions and the growth of the size of nanoclusters, suppressing the paramagnetic and ferromagnetic responses, respectively. 
Figure~\ref{CoAgg} shows schematic images of the Co-ion distribution in Zn$_{1-x}$Co$_x$O. 
According to this picture, in the ferromagnetic hetero-epitaxial Zn$_{1-x}$Co$_{x}$O thin film previously measured by XMCD \cite{PRB_05_Kobayashi}, there were possibly few paramagnetic isolated Co ions while there were smaller nanoclusters than the present homo-epitaxial films [see Fig.~\ref{CoAgg}(c)]. 
Because the present homo-epitaxial films are expected to have higher crystal quality than the previous hetero-epitaxial ones, the nanocluster formation (or the aggregation of the Co ions) may be prevented compared to the hetero-epitaxial Zn$_{1-x}$Co$_{x}$O film. 
Since the annealing suppresses the magnetization and the $n$ is probably unchanged by the annealing, the suppression is attributed not to be originated from the mechanisms related to oxygen vacancy such as bound magnetic polaron \cite{PRL_02_Kaminski} and donor impurity band exchange \cite{NatMater_05_Coey}.

It should be noted that the model based on the spinodal decomposition cannot explain the carrier-related ferromagnetic properties of Zn$_{1-x}$Co$_{x}$O such as the anomalous Hall effects \cite{APL_08_Hsu, JAP_08_Yang}. 
It is possible that there are two origins of ferromagnetism in Zn$_{1-x}$Co$_{x}$O, that is, the formation of antiferromagnetic Co-rich wurtzite (Zn,Co)O nanoclusters and the electron carrier-mediated ferromagnetic interaction. 
In typical ferromagnetic DMS Ga$_{1-x}$Mn$_{x}$As, the origin of ferromagnetism is considered to be due to $p$-$d$ exchange interaction for high carrier concentration and formation of bound magnetic polarons for low carrier concentration. 
Considering that the value of $M_s$ of Zn$_{1-x}$Co$_{x}$O ($< 2$ $\mu_\mathrm{B}$/Co) is smaller than the full moment of Co$^{2+}$ (3 $\mu_\mathrm{B}$/Co), the antiferromagnetic interaction between neighboring Co ions and the carrier-mediated ferromagnetic interaction between distant Co ions may simultaneously affect the magnetic properties of Zn$_{1-x}$Co$_{x}$O. 
Based on the correlation between $M_s$ and $n$ shown in Fig.~\ref{MNvsP}, it is probable that in Zn$_{1-x}$Co$_x$O, the spinodal decomposition is predominant in low $n$ region ($n \lesssim 10^{19}$ cm$^{-3}$), while carrier-induced ferromagnetic interaction is predominant rather than the antiferromagnetic interaction in high $n$ region ($n \gtrsim 10^{19}$ cm$^{-3}$). 
It follows that electron carrier-induced ferromagnetism in Zn$_{1-x}$Co$_x$O possibly needs a highly crystallized Zn$_{1-x}$Co$_{x}$O thin film with high electron carrier concentration $n \gtrsim 10^{19}$ cm$^{-3}$.

\section{conclusion}
We have performed XMCD measurements on homo-epitaxial Zn$_{1-x}$Co$_{x}$O thin films which were prepared with various relatively high oxygen pressures and showed both ferromagnetic and paramagnetic behaviors. The line shapes of the Co $2p$ XAS and XMCD spectra are the same in all the samples, and indicate that the Co ions substitute for the Zn sites. The $H$ and $T$ dependences of Co $2p$ XMCD suggest that paramagnetic behavior is dominant in the samples. 
Analyses using the Curie-Weiss law have revealed that the Curie constant of the samples are close to the value of Co$^{2+}$ and the Weiss temperature are negative, indicating that the paramagnetic Co$^{2+}$ ions interact antiferromagnetically with each other. 
The saturation magnetizations estimated using the XMCD sum rules were almost the same as the magnetization measurements, suggesting that the ferromagnetic responses of the thin films are predominantly caused by magnetic moment of the doped Co ions and confirming that we have probed the bulk magnetic properties. 
The paramagnetic susceptibility and saturation magnetization decreased by high-vacuum annealing. 
The suppression of the paramagnetism by annealing can be explained by thermal diffusion and aggregation of Co ions leading to spinodal decomposition and the antiferromagnetic interaction between the Co ions. Therefore, we propose that the formation of Co-rich nanoclusters coexists and/or competes with carrier-mediated ferromagnetic interaction in Zn$_{1-x}$Co$_{x}$O depending on electron carrier concentration. 
Further systematic XMCD measurements on Zn$_{1-x}$Co$_{x}$O with controlled electron-carrier and/or oxygen-vacancy concentrations will provide us with understandings of the carrier-induced ferromagnetism in Zn$_{1-x}$Co$_{x}$O.

\section*{Acknowledgments}
This work was supported by a Grant-in-Aid for Scientific Research in Priority Area ``Creation and Control of Spin Current'' (19048012) from MEXT, Japan. 
MK acknowledges support from the Japan Society for the Promotion of Science for Young Scientists.

\begin{table}[!p]
\caption{Results of the Curie-Weiss analysis for the XMCD spectra of the Zn$_{1-x}$Co$_{x}$O thin films. Here, $C$ is the Curie constant, $\theta$ is the Weiss temperature, $M_s^{\mathrm{XMCD}}$ and $M_s^{\mathrm{SQUID}}$ are the saturation magnetizations estimated from the XMCD sum rules and measured by the SQUID, respectively. }
\begin{tabular}{lc|cccc}
\hline
\hline
Samples &  \textbf{}  &  $C$ ($\mu_\mathrm{B}$K/T) & $\theta$ (K) & $M_s^{\mathrm{XMCD}}$ ($10^{-2} \mu_\mathrm{B}$/Co) & $M_s^{\mathrm{SQUID}}$ ($10^{-2} \mu_\mathrm{B}$/Co) \\
\hline
$x$=0.07 & $P_{\mathrm{O}2}$=10$^{-4}$ Pa & $3.8\pm 0.5$ & $-37.17\pm 5$ & $0.27 \pm 2.0$ & - \\
$x$=0.10 & $P_{\mathrm{O}2}$=10$^{-2}$ Pa & $1.9\pm 0.5$ & $-12.3\pm 5$ & $3.4\pm 0.8$ & $2.5\pm 0.3$ \\
\textbf{} & $P_{\mathrm{O}2}$=10$^{-3}$ Pa & $3.0\pm 0.5$ & $-37.0\pm 5$ & $1.6\pm 0.8$ & $1.4\pm 0.3$ \\
\textbf{} & $P_{\mathrm{O}2}$=10$^{-4}$ Pa & $3.8\pm 0.5$ & $-8.7\pm 5$ & $1.7\pm 0.8$ & - \\
\hline
\hline
\end{tabular}
\label{CW_analysis}
\end{table}

\begin{table}[!p]
\caption{Results of the Curie-Weiss analysis for the XMCD spectra of the as-grown and annealed Zn$_{0.9}$Co$_{0.1}$O thin films deposited at $P_\mathrm{O2}=10^{-2}$ Pa. Here, $\chi$ is the paramagnetic susceptibility. }
\begin{tabular}{c|ccc}
\hline
\hline
\textbf{} & $\chi$ ($\mu_\mathrm{B}$/Co T) & $M_s^{\mathrm{XMCD}}$ ($10^{-2} \mu_\mathrm{B}$/Co) & $M_s^{\mathrm{SQUID}}$ ($10^{-2} \mu_\mathrm{B}$/Co) \\
\hline
As-grown & 0.059 & $3.4\pm 0.8$ & $2.5\pm 0.3$ \\
Annealed & 0.031 & $1.0\pm 0.8$ & $1.7\pm 0.3$ \\
\hline
\hline
\end{tabular}
\label{EffectsOfAnnealing}
\end{table}

\begin{thebibliography}{10}

\bibitem{Science_98_Ohno}
H. Ohno, Science {\bf 281},  951  (1998).

\bibitem{JSAP_02_Ohno}
H. Ohno, F. Matsukura, and Y. Ohno, JSAP International {\bf 5},  4  (2002).

\bibitem{RMP_06_Jungwirth}
T. Jungwirth, J. Sinova, J. Ma\u{s}ek, J. Ku\u{c}era, and A.~H. MacDonald, Rev.
  Mod. Phys. {\bf 78},  809  (2006).

\bibitem{Nature_99_Ohno}
Y. Ohno, D.~K. Young, B. Beschoten, F. Matsukura, H. Ohno, and D.~D. Awschalom,
  Nature {\bf 402},  790  (1999).

\bibitem{Nature_00_Ohno}
H. Ohno, D. Chiba, F. Matsukura, T. Omiya, E. Abe, T. Dietl, Y. Ohno, and K.
  Ohtani, Nature {\bf 408},  944  (2000).

\bibitem{Science_03_Chiba}
D. Chiba, M. Yamanouchi, Matsukura, and H. Ohno, Science {\bf 301},  943
  (2003).

\bibitem{Nature_04_Yamanouchi}
M. Yamanouchi, D. Chiba, F. Matsukura, and H. Ohno, Nature {\bf 428},  539
  (2004).

\bibitem{Science_00_Dietl}
T. Dietl, H. Ohno, F. Matsukura, J. Cibert, and D. Ferrand, Science {\bf 287},
  1019  (2000).

\bibitem{SST_02_Sato}
K. Sato and H. Katayama-Yoshida, Semicond. Sci. Technol. {\bf 17},  367
  (2002).

\bibitem{SST_04_Pearton}
S.~J. Pearton, W.~H. Heo, M. Ivill, D.~P. Norton, and T. Steiner, Semicond.
  Sci. Technol. {\bf 19},  R59  (2004).

\bibitem{APL_01_Ueda}
K. Ueda, H. Tabata, and T. Kawai, Appl. Phys. Lett. {\bf 79},  988  (2001).

\bibitem{AdvMater_04_Schwartz}
D.~A. Schwartz and D.~R. Gamelin, Adv. Mater. {\bf 16},  2115  (2004).

\bibitem{PRL_06_Neal}
J.~R. Neal, A.~J. Behan, R.~M. Ibrahim, H.~J. Blythe, M. Ziese, A.~M. Fox, and
  G.~A. Gehring, Phys. Rev. Lett. {\bf 96},  197208  (2006).

\bibitem{APL_08_Hsu}
H.~S. Hsu, C.~P. Lin, H. Chou, and J.~C.~A. Huang, Appl. Phys. Lett. {\bf 93},
  142507  (2008).

\bibitem{JAP_08_Yang}
Z. Yang, M. Biasini, W.~P. Beyermann, M.~B. Katz, O.~K. Ezekoye, X.~Q. Pan, Y.
  Pu, J. Shi, Z. Zuo, and J.~L. Lin, J. Appl. Phys. {\bf 104},  113712  (2008).

\bibitem{PRB_07_Matsui}
H. Matsui and H. Tabata, Phys. Rev. B {\bf 75},  014438  (2007).

\bibitem{PRB_05_Kobayashi}
M. Kobayashi, Y. Ishida, J.~I. Hwang, T. Mizokawa, A. Fujimori, K. Mamiya, J.
  Okamoto, Y. Takeda, T. Okane, Y. Saitoh, Y. Muramatsu, A. Tanaka, H. Saeki,
  H. Tabata, and T. Kawai, Phys. Rev. B {\bf 72},  201201(R)  (2005).

\bibitem{PRL_07_Sati}
P. Sati, C. Deparis, C. Morhain, S. Sch{\"{a}}fer, and A. Stepanov, Phys. Rev.
  Lett. {\bf 98},  137204  (2007).

\bibitem{PRB_07_Barla}
A. Barla, G. Schmerber, E. Beaurepaire, A. Dinia, H. Bieber, S. Colis, F.
  Scheurer, J.-P. Kappler, P. Imperia, F. Nolting, F. Wilhelm, A. Rogalev, D.
  M{\"{u}}ller, and J.~J. Grob, Phys. Rev. B {\bf 76},  125201  (2007).

\bibitem{APL_08_Rode}
K. Rode, R. Mattana, A. Anane, V. Cros, E. Jacquet, J.-P. Contour, F. Petroff,
  A. Fert, M.-A. Arrio, P. Sainctavit, P. Bencok, F. Wilhelm, N.~B. Brookes,
  and A. Rogalev, Appl. Phys. Lett. {\bf 92},  012509  (2008).

\bibitem{PRL_06_Kittilstved}
K.~R. Kittilstved, D.~A. Schwartz, A.~C. Tuan, S.~M. Heald, S.~A. Chambers, and
  D.~R. Gamelin, Phys. Rev. Lett. {\bf 97},  037203  (2006).

\bibitem{AdvMater_07_MacManus-Driscoll}
J.~L. MacManus-Driscoll, N. Khare, Y. Lin, and M.~E. Vickers, Adv. Mater. {\bf
  19},  2925  (2007).

\bibitem{Nanotech_06_Wu}
Z.~Y. Wu, F.~R. Chen, J.~J. Kai, W.~B. Jian, and J.~J. Lin, Nanotechnology {\bf
  17},  5511  (2006).

\bibitem{JPCM_07_Sudakar}
C. Sudakar, P. Kharel, G. Lawes, R. Suryanarayanan, R. Naik, and V.~M. Naik, J.
  Phys.: Condens. Mat. {\bf 19},  026212  (2007).

\bibitem{APL_06_Hsu}
H.~S. Hsu, J.~C.~A. Huang, Y.~H. Huang, Y.~F. Liao, M.~Z. Lin, C.~H. Lee, J.~F.
  Lee, S.~F. Chen, L.~Y. Lai, and C.~P. Liu, Appl. Phys. Lett. {\bf 88},
  242507  (2006).

\bibitem{APL_07_Zhao}
Z.~W. Zhao, B.~K. Tay, J.~S. Chen, J.~F. Hu, B.~C. Lim, and G.~P. Li, Appl.
  Phys. Lett. {\bf 90},  152502  (2007).

\bibitem{PRB_06_Sundaresan}
A. Sundaresan, R. Bhargavi, N. Rangarajan, U. Siddesh, and C.~N.~R. Rao, Phys.
  Rev. B {\bf 74},  161306(R)  (2006).

\bibitem{MMM_07_Hong}
N.~H. Hong, J. Sakai, and F. Gervais, J. Magn. Magn. Mater. {\bf 316},  214
  (2007).

\bibitem{JPCM_08_Ghoshal}
S. Ghoshal and P.~S.~A. Kumar, J. Phys.: Condens. Mat. {\bf 20},  192201
  (2008).

\bibitem{JSR_98_Yokoya}
A. Yokoya, T. Sekiguchi, Y. Saitoh, T. Okane, T. Nakatani, T. Shimada, H.
  Kobayashi, M. Takao, Y. Teraoka, Y. Hayashi, S. Sasaki, Y. Miyahira, T.
  Harami, and T.~A. Sasaki, J. Synchrotron Rad. {\bf 5},  10  (1998).

\bibitem{AIP_04_Okamoto}
J. Okamoto, K. Mamiya, S.-I. Fujimori, T. Okane, Y. Saitoh, Y. Muramatsu, A.
  Fujimori, S. Ishikawa, and M. Takano, AIP Conf. Proc. {\bf 705},  1110
  (2004).

\bibitem{PRL_92_Thole}
B.~T. Thole, P. Carra, F. Sette, and G. van~der Laan, Phys. Rev. Lett. {\bf
  68},  1943  (1992).

\bibitem{PRL_93_Carra}
P. Carra, B.~T. Thole, M. Altarelli, and X. Wang, Phys. Rev. Lett. {\bf 70},
  694  (1993).

\bibitem{PRB_07_Dietl}
T. Dietl, T. Andrearczyk, A. Lipi\'{n}ska, M. Kiecana, M. Tay, and Y. Wu, Phys.
  Rev. B {\bf 76},  155312  (2007).

\bibitem{PRL_97_Kodama}
R.~H. Kodama, S.~A. Makhlouf, and A.~E. Berkowitz, Phys. Rev. Lett. {\bf 79},
  1393  (1997).

\bibitem{PRB_05_Winkler}
E. Winkler, R.~D. Zysler, M.~Vasquez Mansilla, and D. Fiorani, Phys. Rev. B {\bf
  72},  132409  (2005).

\bibitem{PRB_05_Lawes}
G. Lawes, A.~S. Risbud, A.~P. Ramirez, and R. Seshadri, Phys. Rev. B {\bf 71},
  045201  (2005).

\bibitem{PRL_02_Kaminski}
A. Kaminski and S.~Das Sarma, Phys. Rev. Lett. {\bf 88},  247202  (2002).

\bibitem{NatMater_05_Coey}
J.~M.~D. Coey, M. Venkatesan, and C.~B. Fitzgerald, Nat. Mater. {\bf 4},  173
  (2005).

\end{thebibliography}
\end{document}